\documentstyle{article}
\renewcommand{\theequation}{\thesection . \arabic{equation} }
\title{\bf On the black-hole kink}
\author{ Pedro F. Gonz\'alez-D\'{\i}az$^{*}$ .\\
Centro de F¡sica "Miguel Catal n",\\
Instituto de Matem\'aticas y F\'{\i}sica Fundamental,\\
Consejo Superior de Investigaciones Cient¡ficas,\\
Serrano 121, 28006 Madrid (SPAIN)\\
}
\date{May 16, 1996}
\begin{document}
\maketitle
\large
\setlength{\baselineskip}{0.9cm}

\begin{center}
{\bf Abstract}
\end{center}

By allowing the light cones to tip over on hypersurfaces
according to the conservation laws of an one-kink in static,
Schwarzschild black hole metric, we
show that in the quantum regime
there also exist instantons whose finite imaginary action
gives the probability of occurrence of the kink metric
corresponding to single
chargeless, nonrotating black holes taking place in pairs,
the holes of each pair being
joined on an interior surface, beyond the
horizon.

\vspace{1cm}

\noindent PACS number(s): 04.70.Bw, 04.70.Dy, 04.60.Gw

\vspace{5cm}

\noindent $^{*}$E-mail: iodpf21@cc.csic.es

\pagebreak

\renewcommand{\theequation}{\arabic{section}.\arabic{equation}}

\section{\bf Introduction}
\setcounter{equation}{0}

The idea that we shall explore in this paper is based on looking at
the spacetime metric of a (D-1)-wormhole as the metric that results
on the constant-time hypersurfaces corresponding to purely future
directed or purely past directed light-cone orientations of a
D-dimensional black-hole spacetime where we allow all possible
light cone orientations compatible with the existence of a
gravitational kink$^{1}$. We shall restrict to the physically most
interesting example with D=4, which associated a Schwarzschild
black hole to a three-dimensional wormhole.

We first briefly review the general topological concept
of a kink and its associated topological charge.
Let $({\bf M},g_{ab})$ be a given D-dimensional spacetime, with
$g_{ab}$ a Lorentz metric on it. One can always regard $g_{ab}$
as a map from any connected D-1 submanifold $\Sigma\subset {\bf M}$
into a set of timelike directions in ${\bf M}$ $^{2}$. Metric homotopy
can then be classified by the degree of this map. This is
seen by introducing a unit line field $\{n,-n\}$, normal to
$\Sigma$, and a global framing $u_{i}$: $i$=1,2,...,D-1, of
$\Sigma$. A timelike vector ${\bf v}$ can then be written in terms
of the resulting tetrad framing $(n,u_{i})$ as
$v=v^{0}n+v^{i}u_{i}$, such that $\sum_{i}^{D-1}(v^{i})^{2}=1$.
Restricting to time orientable manifolds ${\bf M}$, ${\bf v}$ then
determines a map
\[K: \Sigma\rightarrow S^{D-1}\]
by assigning to each point of $\Sigma$ the direction that
${\bf v}$ points to at that point. This mapping allows a general
definition of kink and kink number. Respect to hypersurface
$\Sigma$, the kink number (or topological charge) of the
Lorentz metric $g_{ab}$ is defined by$^{2}$
\[kink(\Sigma;g_{ab})=deg(K),\]
so this topological charge measures how many times the
light cones rotate all the way around as one moves
along $\Sigma$ $^{3}$.

In the case of an asymptotically flat spacetime the pair
$(\Sigma,g)$ will describe an asymptotically flat kink if
$kink(\Sigma;g)\neq 0$. All of the topological charge of
the kink in the metric $g$ is in this case confined to
some finite compact region$^{3}$. Outside that region all
hypersurfaces $\Sigma$ are everywhere spacelike. For the
case of a spherically symmetric kink, to asymptotic
observers, the compact region containing all of the
topological charge coincides with the interior of either
a black hole when the light cones rotate away from the
observers (positive topological charge), or a white
hole when the asymptotic observers "see" light cones
rotating in the opposite direction, toward them (negative
topological charge).

Topology changes, such as handles or wormholes, can occur
in the compact region supporting the kink, but not outside
it. All topologies are actually allowed to happen in such
a region. Therefore, in the case of spherically symmetric
kinks, the supporting region should be viewed as an
essentially quantum-spacetime construct. This is the
view we shall assume throughout this paper.

\section{\bf The Schwarzschild Kink}

We can take for the static, spherically symmetric metric of a
three-dimensional wormhole
\begin{equation}
ds^{2}=(1-\frac{2M}{r})^{-1}dr^{2}+r^{2}d\Omega_{2}^{2},
\end{equation}
where $d\Omega_{2}^{2}$ is the metric on the unit two-sphere.
Metric (2.1) describes a spacetime which (i) is free from any
curvature singularity at $r=0$, and (ii) possesses an apparent
(horizon) singularity at $r=2M$ that is removable by a suitable
coordinate transformation. The re-definition
\begin{equation}
r=\frac{M}{2}\left(\frac{u}{\mu}+\frac{\mu}{u}\right)^{2},
\end{equation}
where $\mu$ is an arbitrary scale, transforms metric (2.1) into
\begin{equation}
ds^{2}=\frac{M^{2}}{4}\left(\frac{u}{\mu}+\frac{\mu}{u}\right)^{4}\left(\frac{4du^{2}}{u^{2}}+d\Omega_{2}^{2}\right).
\end{equation}
Along the complete $u$-interval, ($\infty$,$\mu$), metric (2.3)
varies from an asymptotic region at $u=\infty$ to a minimum
throat at $u=\mu$ (i.e. at $r=2M$).

On the other hand, metric (2.1) is in fact a constant time
section, $T=t_{0}$, of a Schwarzschild black hole,
\begin{equation}
ds^{2}=-(1-\frac{2M}{r})dT^{2}+(1-\frac{2M}{r})^{-1}dr^{2}+r^{2}d\Omega_{2}^{2}.
\end{equation}
Metric (2.1) can likewise be regarded as being described by
constant Euclidean time $\tau=-iT$ sections of the
Gibbons-Hawking instanton$^{4}$ associated to (2.4). Each of
such three-wormholes would correspond to a given Einstein-Rosen
bridge$^{5}$ on this instaton, so that one of the two halves of
the wormhole should then be described in the unphysical$^{4}$
exterior region created in the Kruskal extension of metric
(2.4). In order to avoid the need of using such an unphysical
region to describe a complete wormhole,
we shall consider that metric (2.1) corresponds to a given fixed
value of time $T$ in the kink extension
of (2.4).

We take for the metric that describes a spherically
symmetric one-kink in four dimensions$^{6}$,
\begin{equation}
ds^{2}=-\cos 2\alpha(dt^{2}-dr^{2})\pm 2\sin 2\alpha
dtdr+r^{2}d\Omega_{2}^{2},
\end{equation}
where $\alpha$ is the angle of tilt of the light cones, and
the choice of sign in the second term depends on whether a
positive (upper sign) or negative (lower sign) topological
charge is being considered. An one-kink is ensured to exist
if $\alpha$ is allowed to monotonously increase from 0 to
$\pi$, starting with $\alpha(0)=0$. Then metric (2.5)
converts into (2.4) if we use the substitution
\begin{equation}
\sin\alpha=\sqrt{\frac{M}{r}}
\end{equation}
and introduce a change of time variable $t+g(r)=T$, with
\begin{equation}
\frac{dg(r)}{dr}=\tan 2\alpha .
\end{equation}
Now, since $\sin\alpha$ cannot exceed unity, it follows that
$\infty\geq r\geq M$, so that $\alpha$ varies only from 0
to $\frac{\pi}{2}$. In order to have a complete one-kink
gravitational defect, we need therefore a second coordinate
patch  to describe the other half of the $\alpha$ inteval,
$\frac{\pi}{2}\leq\alpha\leq\pi$.

The kink metric (2.5), which is defined by coordinates $t$, $r$,
$\theta$,$\phi$ and satisfy (2.6) and (2.7), restricts the
Schwarzschild solution to cover only the region
$\infty\geq r\geq M$. If one wants to extend such a metric
to describe the region beyond $r=M$ as well, two procedures
can in principle be followed: (i) if the compact support of
the kink is assumed to be classical, then one lets $\alpha$
continue to increase as $r$ decreases from $r=M$ until
$\alpha=\pi$ at $r=0$ to produce a manifold which has a
homotopically nontrivial light cone field and one kink.
This procedure makes metric (2.5) and definitions (2.6) and (2.7)
to hold asymptotically only, and since, classically, one
should assume a continuous distribution of matter in the
kink support, the momentum-energy tensor can be chosen to
satisfy reasonable physical conditions such as the weak
energy condition$^{6}$. (ii) The second procedure
can apply when
one assumes the black hole interior (i.e. the supporting
compact region of the kink) to be governed by quantum mechanics.
The simplest quantum condition to be satisfied by the
interior region supporting a black- or white-kink arises from
imposing metric (2.5) to hold along the radial
coordinate interval of the kink, i.e.: $\infty\geq r\geq M$,
rather than asymptotically only. Actually, in this case,
the kink geometry
should hold in the two coordinate patches which we need to
describe the complete one-kink gravitational defect. The
need for a second coordinate patch can most clearly be seen
by introducing the new time coordinate
\begin{equation}
\bar{t}=t+h(r),
\end{equation}
which transforms metric (2.5) into the standard metric$^{6}$
\begin{equation}
ds^{2}=-\cos 2\alpha d\bar{t}^{2}\mp 2kd\bar{t}dr+r^{2}d\Omega^{2}_{2},
\end{equation}
provided
\begin{equation}
\frac{dh(r)}{dr}=\frac{dg(r)}{dr}-\frac{k}{\cos 2\alpha},
\end{equation}
with $k=\pm 1$ and the choice of sign in the second term
of (2.9) again depending on whether a positive (upper sign)
or negative (lower sign) topological charge is considered.
The choice of sign in (2.10) is adopted for
the following reason. The zeros of the denominator of
$dh/dr=(\sin 2\alpha\mp 1)/\cos 2\alpha$ correspond to
the two horizons where $r=2M$, one per patch. For the first
patch, the horizon occurs at $\alpha=\frac{\pi}{4}$ and
therefore the upper sign is selected so that both $dh/dr$
and $h$ remain well defined and hence the kink is not lost
in the transformation from (2.5) to (2.9). For the second
patch the horizon occurs at $\alpha=\frac{3\pi}{4}$ and
therefore the lower sign in (2.10)
is selected. $k=+1$ will then
correspond to the first coordinate patch and $k=-1$ to
the second one.

Metric (2.9) can be transformed directly into the
Schwarzschild metric (2.4) if we use (2.6) and the new
coordinate transformation
\begin{equation}
\bar{t}=T-f(r),
\end{equation}
where
\begin{equation}
\frac{df(r)}{dr}=\frac{k}{\cos 2\alpha}.
\end{equation}

We impose then the standard kink metric (2.9) to hold along
$\infty\geq r\geq M$ on the two patches $k=\pm 1$, similarly
to as it has been made in the de Sitter kink$^{7}$. It can
be seen that this condition respects conservation of
energy-momentum tensor only if we assume an energy spectrum
$\frac{kn}{M}$, $n=0,1,2,...$ to hold in the compact, internal
region supporting the kink$^{3}$:
along the radial coordinate
interval $2M\geq r>M$ of the first patch there would be
no spherical surface with nonzero energy and therefore this
interval does not contribute the stress tensor $T_{\mu\nu}$;
as one gets at $r=M$ on the first patch it would appear
a "delta-function-like" concentration of positive energy on
that surface corresponding to the quantum level $n=1$. This
would at first glance blatantly violate energy-momentum
conservation. However, the continuity of the angle of
tilt $\alpha$ at $\frac{\pi}{2}$ implies that the two
coordinate patches are identified at exactly the surfaces
$r=M$. Thus, since there would be an identical "delta-function-like"
concentration of negative energy-momentum at
$n=1$ on $r=M$ in the
second patch, the total stress tensor $T_{\mu\nu}$ will also be
zero at the minimal surface $r=M$. Thus, although the
Birkhoff's theorem ensures$^{8}$ the usual Schwarzschild metric
as the unique spherically symmetric solution to the
four-dimensional vacuum Einstein equation, the violation
of this classical result the way we have shown above implies
an allowed {\it quantized} extension from it because this
extension entails no violation of energy-momentum conservation at any
interior spacelike hypersurface.

This result should be interpreted as follows. All what is
left at length scales equal or smaller than the minimum
size of the bridge (i.e. for $n\geq 1$, $r\leq M$) is some
sort of quantized "closed" baby universe with maximum size
$M$, whose zero total energy may be regarded as the sum of
the opposite-sign eigenenergies of two otherwise identical
harmonic oscillators with the zero-point energy substracted.
The positive energy oscillator would play the role of the
matter field part of a constrined Hamiltonian, $H=0$, and
the negative energy oscillator would behave like though it
were the gravitational part of this Hamiltonian
constraint. On the
other hand, to an asymptotic observer in either patch, the above
quantized kink geometry would look like that of a black hole if the
topological charge is positive, and like that of a white hole if
the topological charge is negative. In the latter case,
to the asymptotic observer there would actually be
a topological change by which
an asymptotically flat space converts into asymptotically
flat space plus a baby universe being branched off from it.
Now, since from a quantum-mechanical standpoint white and
black holes with the same mass are physically indistinguishable$^{9}$,
it follows that to asymptotic observers
the asymptotically flat space of black holes is
physically indistinguishable from
asymptotically flat space plus a baby
universe, with such a baby universe living outside the
realm of the two coordinate patches where the kink is
defined, in the inaccessible region between $r=M$ and
$r=0$.

Metric (2.9) contains still the geodesic incompleteness at
$r=2M$ of metric (2.4). This incompleteness can be removed by
the use of Kruskal technique. Thus, introducing the metric
\begin{equation}
ds^{2}=-F(U,V)dUdV+r^{2}d\Omega_{2}^{2},
\end{equation}
in which
\begin{equation}
F=\frac{4M\cos 2\alpha}{\beta}\exp\left(-2\beta
k\int^{2}_{\infty/M}\frac{dr}{\cos 2\alpha}\right),
\end{equation}
\begin{equation}
U=\mp e^{\beta\bar{t}}\exp\left(2\beta
k\int^{r}_{\infty/M}\frac{dr}{\cos 2\alpha}\right),
\end{equation}
\begin{equation}
V=\mp\frac{1}{2\beta M}e^{-\beta\bar{t}},
\end{equation}
where $\beta$ is an adjustable parameter which will be chosen
so that the unphysical singularity at $r=2M$ is removed,
and the lower integration limit $\infty/M$ refers to the
choices $r=\infty$ and $r=M$, depending on whether the
first or second patch is being considered. Using (2.6)
we obtain from (2.14)
\[F=4M\left(\frac{1-\frac{2M}{r}}{\beta}\right)\left(\frac{r}{M}\left(\frac{2M}{r}-1\right)\right)^{-4\beta
kr}.\]
This expression would actually have some constant term
coming from the lower integration limits $\infty/M$. We
have omitted at the moment such a term because it is
canceled by the similar constant term which appears
in the Kruskal coordinate $U$ when forming the Kruskal metric
from (2.13)-(2.16).

Unphysical singularities are then avoided if we choose
\begin{equation}
\beta=\frac{1}{4kM} .
\end{equation}
Whence
\begin{equation}
F=\frac{16kM^{3}}{r}e^{-\frac{r}{2M}},
\end{equation}
\begin{equation}
U=\mp e^{\frac{\bar{t}}{4kM}}e^{\frac{r}{2M}}(\frac{2M-r}{M}),
\; \; \; V=\mp\frac{k}{2}e^{-\frac{\bar{t}}{4kM}},
\end{equation}
where$^{1}$
\[\bar{t}=t_{0}-k\int_{\infty/M}\frac{dr}{\cos 2\alpha}\]
\begin{equation}
=\bar{t}_{0}-k\left(r-2M\ln\left(\frac{M}{2M-r}\right)\right),
\end{equation}
with the constant $\bar{t}_{0}$ being obtained from $t_{0}$ after
absorbing the term arising from the lower integration limit
$\infty$ or $M$, depending on whether the first or second patch
is being considered. We finally obtain for the Kruskal metric
of the Schwarzschild kink
\begin{equation}
ds^{2}=-\frac{32kM^{3}}{r}e^{-\frac{r}{2M}}dUdV+r^{2}d\Omega_{2}^{2}.
\end{equation}
Except for the sign parameter $k$, this metric is the same as the
Schwarzschild-Kruskal metric.

Because of continuity of the angle of tilt $\alpha$ at $\frac{\pi}{2}$,
the two coordinate patches can be identified to each other only
on the surfaces at $r=M$ $^{1,10}$. Such an identification should occur
both on the original and the new regions created by the Kruskal
extension, and represents a bridge that connects asymptotically
flat regions of the two coordinate patches. Any $T=$const. section
of this spacetime construct will then describe
halves of a three-dimensional
wormhole whose neck is now at $r=M$, rather than $r=2M$.
One can then describe the two halves of a complete
wormhole just in the physical
original regions of either patch $k=+1$ or patch $k=-1$.

The causal structure of the considered geometry could at first
glance be thought of as being unstable due to mass-inflation
caused by the unavoidable presence of a Cauchy horizon$^{11}$:
because quanta that enter the future event horizon at
arbitrary late time suffers an arbitrarily large blue shift
while propagating parallel to the Cauchy horizon, there will
be in general a mass-inflation singularity along a part of the
horizon in one patch caused by small fluctuations in the other.
However, using the spherical shell approach in the lightlike
limit$^{12}$ where a mass shell is allowed to move toward $r=0$
in the field of an interior mass distribution, it can be shown
that our kink model with quantized support prevents the
occurrence of any mass-inflation singularity. In fact, any
interior energy fluctuation in one patch is necessarily
sign-reversed to the energy of the imploding shell in the
other. Therefore, a mass increase must now occur in the
expanding fluctuation shell, rather than in the imploding
shell, and the mass variation of these two shells is
nonsingular everywhere for $r\geq M$, even at the collision
radius where one would expect the mass singularity to
occur. We actually expect that, at that radius, imploding
and expanding gravitational masses are both finite with
half and twice their respective asymptotic values.

\section{\bf Euclidean formalism}
\setcounter{equation}{0}

The Euclidean section of the Schwarzschild solution is asymptotically
flat and nonsingular because it does not contain any points
with $r<2M$. Thus, the curvature singularity does not lie
on the Euclidean section. Here I shall consider
the instantons that can be associated with the black hole kinks,
and show that their Euclidean sections can be extended beyond
the horizon down to the surface $r=M$.

The Euclidean continuation of the metrics which contain one kink
should be obtained by putting
\begin{equation}
\bar{t}=i\bar{\tau}.
\end{equation}
Using (2.6) and (2.7) we then have
\begin{equation}
d\bar{\tau}=-idT+\frac{ik}{\cos 2\alpha}dr.
\end{equation}
This Euclidean
continuation would give rise to metrics which are positive
definite if we choose either the usual continuation $T=i\tau$,
for $r\geq 2M$, or the new Euclidean continuation $r=-i\rho$,
$M=-i\mu$, for $r<2M$, where $r$ becomes timelike, and we
transform a space coordinate into a time coordinate. In the
first case, metric (2.9) becomes
\begin{equation}
ds^{2}=\cos 2\alpha d\bar{\tau}^{2}\mp 2ikd\bar{\tau}dr+r^{2}d\Omega_{2}^{2}.
\end{equation}
This corresponds to the usual Euclidean subsection $\infty\geq r\geq2M$,
\begin{equation}
ds^{2}=\left(1-\frac{2M}{r}\right)d\tau^{2}
+\left(1-\frac{2M}{r}\right)^{-1}dr^{2}+r^{2}d\Omega_{2}^{2},
\end{equation}
and can be maximally-extended to the Kruskal metric
\begin{equation}
ds^{2}=-\frac{32M^{3}k}{r}e^{-\frac{r}{2M}}d\tilde{U}d\tilde{V}
+r^{2}d\Omega_{2}^{2},
\end{equation}
where
\begin{equation}
\tilde{U}=\mp e^{\frac{i\tau}{4kM}}e^{\frac{r}{2M}}\left(\frac{2M-r}{M}\right),
\;\;\; \tilde{V}=\mp\frac{k}{2}e^{-\frac{i\tau}{4kM}}.
\end{equation}

In order for the new continuation $r=-i\rho$, $M=-i\mu$ to give rise
to a metric which is positive definite for $r<2M$, we would also
continue the angular polar coordinates such that $\theta=-i\Theta$,
$\phi=\phi$. With this choice we then had for the orthogonal
coordinates
\[x=-X=-\rho\sinh\Theta\cos\phi, \;\;
y=-Y=-\rho\sinh\Theta\sin\phi,\]
\[z=-iZ=-i\rho\cosh\Theta,\]
so that $r=\sqrt{(x^{2}+y^{2}+z^{2})}=\pm i\rho$ and $\mid Z\mid\geq\rho$.
Therefore, we in fact have $\phi=\arctan\frac{y}{x}=\arctan\frac{Y}{X}$,
and $\theta=\arccos\frac{z}{r}=\pm i\Theta$,
with $\Theta=\cosh^{-1}\frac{Z}{\rho}$. Hence, the metric on the unit
two-sphere $d\Omega_{2}^{2}$ should transform as

\[d\Omega_{2}=\pm id\omega_{2}=i(d\Theta^{2}+\sinh^{2}\Theta d\phi^{2})^{\frac{1}{2}}.\]
The choice of the minus sign for the Euclidean continuation of both
the radial coordinate $r$ and the polar angle $\theta$ would allow us
to have the same action continuation as that corresponding to the
continuation $T=i\tau$; i.e. $S=iI$, where $S$ and $I$ are the
Lorentzian and Euclidean action, respectively, since the scalar
curvature transforms as $R(r)=-R(\rho)$ under continuation
$r=-i\rho$.

Thus, for the continuation $r=-i\rho$, $M=-i\mu$, $d\Omega_{2}
=\pm id\omega_{2}$ for $r<2M$, metric (2.9) becomes
\begin{equation}
ds^{2}=\cos 2\alpha d\bar{\tau}^{2}\pm 2kd\bar{\tau}d\rho+\rho^{2}d\omega_{2}^{2},
\end{equation}
which corresponds to the new Euclidean subsection $2M>r>M$, with
positive definite metric
\begin{equation}
ds^{2}=(\frac{2\mu}{\rho}-1)dT^{2}+(\frac{2\mu}{\rho}-1)^{-1}d\rho^{2}
+\rho^{2}d\omega_{2}^{2},
\end{equation}
and can be maximally-extended to the Kruskal metric
\begin{equation}
ds^{2}=+\frac{32\mu^{3}k}{\rho}e^{-\frac{\rho}{2\mu}}d\hat{U}d\hat{V}
+\rho^{2}d\omega_{2}^{2},
\end{equation}
where in this case
\begin{equation}
\hat{U}=\mp e^{\frac{\tau}{4k\mu}}e^{\frac{\rho}{2\mu}}\left(\frac{2\mu-\rho}{\mu}\right),
\;\;\; \hat{V}=\mp\frac{k}{2}e^{-\frac{\tau}{4k\mu}}.
\end{equation}

Using a positive definite metric such as (3.8) leads however
to the problem that the azimuthal angle $\theta$ is a periodic
variable only outside the horizon. In this case the transverse
two-manifold, which is a two-sphere of positive scalar curvature
outside the Euclidean horizon, becomes a hyperbolic plane of
negative scalar curvature inside the horizon and any boundary
at finite geodesic distance inside the horizon is no longer
compact. In particular, the boundary at $r=M$ would then have
the noncompact topology $S^{1}\times R^{2}$, with $S^{1}$
corresponding to time $T$. One cannot identify this geometry
at $\rho=2\mu$ with the geometry at $r=2M$ corresponding to
the compact topology $S^{1}\times S^{2}$ of the Gibbons-Hawking
instanton$^{4}$. Nevertheless, avoidance of the spacetime
singularities in the calculation of the black hole action
does not actually require having a positive definite metric
in our spacetime kinks. Indeed, the "tachyonic continuation"
of the signature + + - - (which is - - + +) that
corresponds to a real azimuthal periodic variable $\theta$
also inside the horizon and implies the same metrics as
(3.7), (3.8) and (3.9) but with the sign for the polar
coordinate terms reversed, can not only avoid singularities
but erase them even at $r=0$. In order to see this, let us
consider the new variables $y+z=U$ and $y-z=V$ in the
Kruskal metric (2.21) which then becomes
\begin{equation}
ds^{2}=-\frac{32kM^{3}}{r}e^{-\frac{r}{2M}}(dy^{2}-dz^{2})+r^{2}d\Omega_{2}^{2},
\end{equation}
with
\begin{equation}
y^{2}-z^{2}=ke^{\frac{r}{2M}}\left(1-\frac{r}{2M}\right)
\end{equation}
\begin{equation}
\frac{y+z}{y-z}=ke^{\frac{\bar{t}}{2kM}}e^{\frac{r}{2M}}\left(\frac{r}{2M}-1\right).
\end{equation}
The singularity at $r=0$ lies on the surfaces $y^{2}-z^{2}=k$.
This singularity
can be avoided by defining either a new coordinate
$\zeta=iy$ or a new coordinate $\xi=iz$. For the first choice
the metric takes the Euclidean form
\begin{equation}
ds^{2}=\frac{32kM^{3}}{r}e^{-\frac{r}{2M}}(dz^{2}+d\zeta^{2})+r^{2}d\Omega_{2}^{2},
\end{equation}
which is positive definite in the patch $k=+1$ and
has in fact signature
- - + + in the patch $k=-1$. The radial coordinate is then defined
by
\begin{equation}
z^{2}+\zeta^{2}=ke^{\frac{r}{2M}}\left(\frac{r}{2M}-1\right).
\end{equation}

On the section on which $z$ and $\zeta$ are both real (the usual
Euclidean section for patch $k=+1$) $\frac{r}{2M}$ will be real
and greater or equal to 1 on patch $k=+1$, and
$\frac{1}{2}\leq\frac{r}{2M}\leq 1$ on patch $k=-1$, the lower
limit $\frac{1}{2}$ being imposed by the continuity of the kink
at $\alpha=\frac{\pi}{2}$. Define the imaginary time by
$T=i\tau$. This continuation leaves invariant the form of
the metric (3.14) and is therefore compatible with the
coordinate transformation $\zeta=iy$. Then, from (2.20) and
(3.13) we obtain
\begin{equation}
z-i\zeta=\pm\left(z^{2}+\zeta^{2}\right)^{\frac{1}{2}}e^{\frac{i\tau}{4kM}}.
\end{equation}
It follows that for this time continuation $\tau$ is periodic
with period $8\pi kM$. On this nonsingular Euclidean section,
$\tau$ has then the character of an angular coordinate which
rotates about the "axis" $r=2M$ clockwise in patch $k=+1$,
and anticlockwise about the "axis" $r=0$ in patch $k=-1$.
Any boundary $\partial M_{k}$ in this Euclidean section has
topology $S^{1}\times S^{2}$ and so is compact in both
coordinate patches. Since the scalar curvature $R$ vanishes,
the action can be written only in terms of the surface
integrals corresponding to the fixed boundaries. This action
can be written
\begin{equation}
I_{k}=\frac{1}{8\pi}\int_{\partial M_{k}}d^{3}xK_{k},
\end{equation}
where $K_{k}=K-\frac{1}{2}(1+k)K^{0}$, $K$ being the trace
of the second fundamental form of the boundary, and $K^{0}$
the trace of the second fundamental form of the boundary
imbedded in flat space. This action was evaluated$^{2}$ in
the case of the positive definite metric which corresponds
to $k=+1$. It is $I_{+1}(M)=4\pi iM^{2}$. In the case $k=-1$,
fixing the boundary at the surface $r=A=M$, we also have
\[I_{-1}(M)=\frac{1}{8\pi}\int_{\partial M_{-1}^{A}}Kd\Sigma\]
\[=-4\pi i(2r-3M)\mid_{r=M}=4\pi iM^{2}.\]

For the second choice of coordinates, $\xi=iz$, metric (3.11)
takes the form
\begin{equation}
ds^{2}=-\frac{32kM^{3}}{r}e^{-\frac{r}{2M}}(dy^{2}+d\xi^{2})+r^{2}d\Omega_{2}^{2},
\end{equation}
which is positive definite in patch $k=-1$ and has again signature
- - + + in patch $k=+1$. The radial coordinate is now defined by
\begin{equation}
y^{2}+\xi^{2}=ke^{\frac{r}{2M}}\left(1-\frac{r}{2M}\right),
\end{equation}
so that on the section on which $y$ and $\xi$ are both real
(the usual Euclidean section for patch $k=-1$) $\frac{r}{2M}$
will be in the interval $\frac{1}{2}\leq\frac{r}{2M}\leq 1$
on patch $k=+1$, and greater or equal to 1 on patch $k=-1$.
We define now the imaginary $r$ and $M$ by
$r=-i\rho$ and $M=-i\mu$, keeping $T$ and the azimuthal coordinate
$\theta$ real. In order for this definition to be compatible with the
coordinate transformation $\xi=iz$, it should leave metric
(3.18) formally unchanged. For this to be accomplished one
must also continue the line element $ds$ itseft, namely
$ds=-id\sigma$, instead of the azimuthal angle $\theta$.
This requirement becomes most natural if we recall that the
interval $ds$ has the same physical dimension as that of $r$
and $M$, and that the "tachyonic" mass $\mu$ should be
associated with an imaginary relativistic interval. Then,
from (2.20) and (3.13) we obtain
\begin{equation}
y-i\xi=\pm(y^{2}+\xi^{2})^{\frac{1}{2}}e^{\frac{iT}{4k\mu}}.
\end{equation}
It is now the Lorentzian time $T$ which becomes periodic
with period $8\pi k\mu$. On this new nonsingular Euclidean
section,
$T$ would have the character of an angular coordinate
which rotates about the "axis" $\rho=0$ clockwise in the
patch $k=+1$, and anticlockwise about the "axis"
$\rho=2\mu$ in the patch $k=-1$. In such a new section, the
action is given by (3.17), where now
$K_{k}=K-\frac{1}{2}(1-k)K^{0}$. On the patch $k=+1$, we have$^{4}$
\[I_{+1}(\mu)=\frac{1}{8\pi}\int_{\partial M_{+1}^{A}} Kd\Sigma \]
\[=4\pi iM(2r-3M)\mid_{r=M}=-4\pi iM^{2}=4\pi i\mu^{2}.\]
In the patch $k=-1$, taking $K^{0}=\frac{2}{r}$ and following
Gibbons and Hawking$^{4}$, we obtain the action $I_{-1}(\mu)$
which turns out to be the same as $I_{+1}(\mu)$.

Thus, on the coordinate patch $k=+1$, the Euclidean continuation
(3.1) of the time coordinate $\bar{t}$ of the kink metric
contains both the continuation for time $T$, $T=i\tau$,
where the apparent singularity at $r=2M$ is like the
irrelevant singularity at the origin of the polar coordinates
provided that $\frac{\tau}{4M}$ is regarded as an angular
variable and is identified with period $2\pi$ $^{4}$, and a new
continuation $r=-i\rho$, which also implies "tachyonic"
continuations $M=-i\mu$ and $ds=-id\sigma$, where the
curvature singularity at $\rho=0$ becomes again like a
harmless polar-coordinate singularity provided that
$\frac{T}{4\mu}$ is regarded as an angular variable and
is identified with period $2\pi$. The transverse two-manifold
is now a compact two-sphere both outside and inside the
Euclidean horizon and any boundaries have compact topology
$S^{1}\times S^{2}$, with $S^{1}$ corresponding to $\tau$
outside the horizon and to $T$ inside the horizon. Since
these topological products are compact, have the same
Euler characteristic and are both orientable, they are
homeomorphic to each other with a continuous mapping
between them. Therefore, one can identify the two
corresponding geometries at the Euclidean horizons ($r=2M$
and $\rho=2\mu$) which, respectively, $\tau$ rotates about
at zero geodesic distance and is the geodesic distance at
which $T$ rotates about $\rho=0$. The Gibbons-Hawking
instanton can then be extended beyond the Euclidean
horizon down to just the boundary surface at $r=M$ ($\rho=\mu$)
where the first and second patches must be somehow joined
onto each other.
The resulting Euclidean section does not
contain any points with $r<M$ and therefore the curvature
singularity is still avoided, as it also is in the baby
universe sector ($\rho < \mu$) due to
the periodic nature of the instantonic
time $T$. The spacetime of the extended instanton covers
the entire domain of the coordinate patch $k=+1$ and that
of the baby universe is outside the two coordinate patches.

Euclidean continuation (3.1) on coordinate patch $k=-1$
leads to the same instantonic sections as for patch $k=+1$,
but now $T=i\tau$ corresponds to the section inside the
horizon $r=2M$ up to $r=M$, and $r=-i\rho$, $M=-i\mu$,
$ds=-id\sigma$ define the section outside the horizon
$\rho=2\mu$, with $\tau$ and $T$ respectively rotating about
$r=0$ and $2\mu$, anticlockwise in both cases. Since the
boundaries at constant radial coordinates on both sides
of the Euclidean horizon have compact topology $S^{1}\times
S^{2}$, the geometries at the Euclidean horizon ($r=2M$ and
$\rho=2\mu$) can also be identified, leading to an instanton
which covers the entire coordinate patch $k=-1$.

On any $\tau -r$ plane in the coordinate patch $k=+1$ we
can define the amplitude
$\langle\tau_{2}\mid\tau_{1}\rangle$ to go from the surface
$\tau_{1}$ to the surface $\tau_{2}$ which is dominated by
the action $I_{1}(M)=4\pi iM^{2}$, corresponding to the
circular sector limited by the times $\tau_{1}$ and $\tau_{2}$
on a circle centered at $r=2M$ with large radius $r_{0}\gg 2M$.
Similarly, on the $t-\rho$ plane in the patch $k=+1$ the amplitude
$\langle t_{2}\mid t_{1}\rangle$ to go from the surface $t_{1}$
to the surface $t_{2}$ is dominated by the action $I_{2}(\mu)=4\pi
i\mu^{2}$ that corresponds to the sector limited by times
$t_{1}$ and $t_{2}$ from $\rho=\mu$ to $\rho=2\mu$ on a circle
centered at $\rho=0$. An asymptotic observer
in patch $k=+1$ would interpret
these results as providing the probability of the occurrence
in the vacuum state of, respectively, a black hole with mass
$M$ or a white hole with mass $\mu$. In the coordinate patch
$k=-1$, the same observer would reach the same interpretation but
for a black hole with mass $\mu$ or a white hole with mass
$M$.

Any constant time section of these instantons would
represent the half of a three-dimensional
wormhole either in the first or the second coordinate patch.
The connection of one such wormhole halves in the first patch
to other wormhole half in the second patch would take place
on an equatorial surface $r=A$ and produce a complete
wormhole with two original asymptotic regions, one in patch
$k=+1$ and the other in patch $k=-1$.

\vspace{1cm}

\noindent {\bf Acknowledgements}

\noindent
This work was supported by DGICYT
under Research Projects No. PB94-0107-A.

\pagebreak

\noindent \section*{References}
\begin{description}

\item [1] P.F. Gonz\'alez-D\'{\i}az, Phys. Rev. D51, 7144
(1995).
\item [2] G.W. Gibbons and S.W. Hawking, Phys. Rev. Lett.
69, 1719 (1992).
\item [3] A. Chamblin, {\it Kinks and singularities},
DAMTP preprint R95/44.
\item [4] S.W. Hawking, in {\it General Relativity: An Einstein
Centenary Survey}, eds. S.W. Hawking and W. Israel (Cambridge
Univ. Press, Cambridge, UK, 1979); G.W. Gibbons and S.W.
Hawking, Phys. Rev. D15, 2752 (1977);
R.M. Wald, {\it General relativity} (The
University Chicago Press, Chicago, USA, 1984).
\item [5] A. Einstein and N. Rosen, Phys. Rev. 48, 73 (1935).
\item [6] D. Finkelstein and C.W. Misner, Ann. Phys. (N.Y.) 6 230
(1959);
D. Finkelstein, in: {\it Directions in General Relativity I},
eds. B.L. Hu, M.P. Ryan Jr. and C.V. Vishveshwara (Cambridge Univ. Press,
Cambridge, 1993).
\item [7] K.A. Dunn, T.A. Harriot and J. G. Williams,
J. Math. Phys. 35, 4145 (1994).
\item [8] S.W. Hawking and G.F.R. Ellis, {\it The
Large Scale Structure of Space-Time} (Cambridge University
Press, Cambridge, UK, 1973).
\item [9] S.W. Hawking, Phys. Rev. D13, 191 (1976);
D14, 2460 (1976).
\item [10] D. Finkelstein and G. McCollum, J. Math. Phys. 16 2250
(1975).
\item [11] E. Poisson and W. Israel, Phys. Rev. D41, 1796
(1990); A. Ori, Phys. Rev. Lett. 67, 789 (1991); 68, 2117
(1992).
\item [12] C. Barrab\`es and W. Israel, Phys. Rev.
D43, 1129 (1991).

\end{description}

\end{document}